\documentstyle[preprint]{jpsj}

\newcommand{\noi}{\noindent}
\newcommand{\non}{\nonumber}

\newcommand{\ii}{{\rm i}}
\newcommand{\kpara}{k_{\parallel}}
\newcommand{\kperp}{k_{\perp}}

\newcommand{\para}{\parallel}
\newcommand{\beq}{\begin{eqnarray}}
\newcommand{\eeq}{\end{eqnarray}}

\newcommand{\be}{\begin{eqnarray}}
\newcommand{\en}{\end{eqnarray}}
\newcommand{\JPSJ}{J. Phys. Soc. Jpn.}
\newcommand{\JLTP}{J. Low Temp. Phys}
\newcommand{\PRL}{ Phys. Rev. Lett.}
\newcommand{\PRB}{ Phys. Rev.{\bf  B}}
\newcommand{\PR}{ Phys. Rev.}
\newcommand{\PROG}{Prog. Theor. Phys.}
\def\runtitle{Crossover from  Antiferromagnetic  Phase to   Fermiology Regime in a Weakly Coupled  Half-Filled Chain System}
\def\runauthor{Jun-ichiro Kishine and Kenji Yonemitsu}
\title{Crossover from  Antiferromagnetic  Phase to   Fermiology Regime in a Weakly Coupled  Half-Filled Chain System}
\author{Jun-ichiro Kishine\thanks{E-mail:kishine@ims.ac.jp} and Kenji Yonemitsu}
\inst
{Department of Theoretical Studies,
Institute for Molecular Science,
Okazaki 444-8585, Japan}
\recdate{today}
\abst
{Effects of the electron-electron umklapp process on dimensional crossovers caused by  
the interchain one-particle hopping, $t_{\perp}$, in  a weakly-coupled half-filled
chain system  have been studied, based on the perturbative renormalization-group approach.
The variance of $t_\perp$ and the   umklapp process under scaling is taken into account.
We found that the intrachain umklapp process causes  a finite crossover value of $t_\perp$, $t_{\perp \rm cr}$.
For $t_{\perp}<t_{\perp \rm c}$, as the temperature decreases, the system undergoes
a crossesover from the Tomonaga-Luttinger (TL) liquid to an incipient one-dimensional Mott insulator   and finally 
makes a phase transition into 
an antiferromagnetic  long-range-ordered phase    at a transition temperature $T_{N}$ which increases with the increasing $t_{\perp}$. 
For $t_{\perp}>t_{\perp \rm c}$, the system undergoes a crossesover from the TL liquid to the Fermiology regime
where the interchain   propagation of a quasi-particle is coherent.
We discuss relevance of the present result to the experimentally suggested, pressure-induced crossover phenomena 
in the quasi-one-dimensional organic systems.}
\kword
{  umklapp scattering,   dimensional crossover, Tomonaga-Luttinger liquid, Mott insulator, SDW, Fermiology,
perturbative renormalization-group, quasi-one-dimensional organic system}

\begin{document}
\maketitle
\baselineskip16pt

In one-dimensional (1D)  systems,  quantum fluctuations prevent the system undergo a phase transition toward a long-range ordered phase and consequently 
only {\it incipient instability} can exist.\cite{Solyom}
When 1D systems are weakly coupled via a one-particle hopping, $t_{\perp}$,  they behave as isolated 1D systems at high temperature,
 $T \stackrel{>}{\scriptstyle \sim}t_{\perp}$.
As the temperature decreases, they undergo either of two dimensional crossovers.\cite{Bourbonnais91}
A {\lq\lq}two-particle crossover{\rq\rq} converts   the incipient instability   to the instability toward a long-range-order, while
 a {\lq\lq}one-particle crossover{\rq\rq} converts the transverse incoherent propagation of a quasi-particle to a coherent one.
In the latter case, below the crossover temperature, the Fermi surface effects become important.
The scaling theory of dimensional crossovers based on the perturbative renormalization-group (PRG) approach\cite{Bourbonnais91}
tells us 
which crossover the system undergoes  essentially depends on the {\it geometry} (chain, ladder...) and the {\it universality} (weak-coupling or strong-coupling) 
of the corresponding isolated system.\cite{JKKY}

In quasi-one-dimensional organic systems, (TMTTF)$_2$X   (X=Br, PF$_6$, SCN...),
     a dimerization  along the conducting stack  makes  the band   effectively half-filled instead of three-quarter filled as expected from the 2:1
 stoichiometry.\cite{Jerome91}
At ambient pressure, (TMTTF)$_2$Br is semiconducting with a shallow minimum in resistivity around $T_{\rho}\sim 100$K 
and undergoes a phase transition at $T_{N}\sim 13$K   to a commensurate spin-density wave (CSDW) phase characterized by the SDW  amplitude
$0.14\mu_{B}$/molecule and the
wave-number (1/2,1/4,0).\cite{Nakamura95}
Recently the irrelevance of  the Fermi surface to a CSDW transition    was reported in  (TMTTF)$_2$SCN\cite{Nakamura97}
which undergoes  a clear metal-insulator
transition accompanying    ordering of the non-centrosymmetric anions (SCN$^-$) at $T=160$K\cite{Coulon82} without any anomaly of magnetic susceptibility and
a phase transition to a CSDW phase occurs at   $T_{N}\sim 7$K with  the SDW  amplitude and the wave-number being identical to those of  (TMTTF)$_2$Br.\cite{Nakamura97}
This fact strongly suggests that the Fermi surface effects are not responsible for the phase transition to the ICSDW phase in these TMTTF-salts.\cite{Nakamura97}

On the other hand, (TMTSF)$_2$PF$_6$ shows  metal-like behavior down to a phase transition at $T_{N}\sim 12$ K to an incommensurate SDW (ICSDW) 
phase characterized by   the SDW amplitude $0.08\mu_{B}$/molecule and the     wave-vector   
$(0.5,0.24\pm0.03,-0.06\pm0.20)$\cite{Takahashi86} which is quite close to
the calculated optimal nesting vector.\cite{Ducasse} 
The incommensurate character  
 is also supported through $\mu$SR studies,\cite{Le91} observations of motional narrowing of the NMR line shape and  non-Ohmic electrical transport caused by a 
pinning   of the ICSDW.\cite{Kang90}
It is well established that the   Fermi surface nesting is responsible for the ICSDW phase of (TMTSF)$_2$PF$_6$.\cite{Yamaji83}
The relevance of the Fermi surface effects is  also supported by the success of Fermiology arguments   on
the field-induced SDW phenomena in the TMTSF-salts.\cite{GorkovLebed,Yamaji85}

Recently pressure-induced crossover phenomena in (TMTTF)$_2$Br were reported, based on     transport and NMR experiments.\cite{Klemme95,Klemme96} 
$dT_{N}/dP>0$ and a commensurate SDW character are observed   at low pressure $P<P_{\rm opt}=5$kbar, whereas $dT_{N}/dP<0$ and an incommensurate  SDW character  
  at low pressure $P>P_{\rm opt}$. The crossover between the CSDW and ICSDW regimes is  coincident with the vanishing of a charge localization 
gap $\Delta_{\rho}$.\cite{Klemme95}
The features of the NMR rate ($T_{1}^{-1}$) in the  ICSDW phase at $P>P_{\rm opt}$ are common to  (TMTSF)$_2$PF$_6$ under ambient pressure.\cite{Klemme96} 

From theoretical viewpoints, the importance of the umklapp process in  (TMTTF)$_2$X and (TMTSF)$_2$X   was first addressed by   Emery {\it et al.}\cite{Emery82} with emphasis 
on the relation between
  the   dimerization and the umklapp process. They treated the dimensionality effects in a qualitative manner by controlling
the  interference between the lowest-order particle-particle and particle-hole fluctuations.

On the other hand, based on the PRG approach, Bourbonnais\cite{Bourbonnais91,Bourbonnais97}   suggested that 
 the two-particle crossover  becomes possible when
the one-particle crossover temperature, $T_{x1}$, is strongly suppressed by  the development of  a charge localization gap
$\Delta_{\rho}$. Then  the  CSDW phase transition is induced by the  interchain coherence of an electron-hole pair 
(a $2k_{F}$ spin correlation). Based on the Stoner criteria (mean field treatment of the interchain coupling), Bourbonnais   
obtained $T_{N}\sim {t_{\perp}^{\ast}}^2/\Delta_{\rho}$ (where
 $t_{\perp}^{\ast}$   denotes a renormalized $t_\perp$) which increases as $\Delta_{\rho}$ decreases under pressure.\cite{Bourbonnais97}
In Bourbonnais's arguments, however,  
  the {\it variance of $t_\perp$ and the   umklapp process under scaling}  
is not explicitly taken into account.

It is also to be noted that based on a mean-field calculation  for a one-dimensional conductor at {\it quarter filling},   Seo and Fukuyama\cite{Seo97} recently
found that the  CSDW phase is accompanied by   $4k_F$ charge ordering   when   the nearest neighbor Coulomb interaction is strong enough. 

In this paper, we develop a scaling theory of the dimensional crossovers in the weakly-coupled half-filled  chain system
{\it by taking account of
 the variance of $t_\perp$ and the   umklapp process under scaling.}
It is pointed out   in our previous paper\cite{JKKY} that the  asymptotic behavior of $t_{\perp}$ plays an essential role in determing which crossover the system undergoes.
We start with the path integral representation of the partition function,
$
Z=\int{\cal D} e^{S},
$
where   the action  consists of four parts,
$
S=S^{(1)}_{\para}+S^{(1)}_{\perp}+S^{(2)}_{\para}+S^{(2)}_{\perp},\label{eqn:action}
$
with 
$S^{(1)}_{\para}$, $S^{(1)}_{\perp}$, $S^{(2)}_{\para}$  and 
$S^{(2)}_{\perp}$ being the actions 
for the {\it intra}chain one-particle hopping, {\it inter}chain 
one-particle hopping, {\it intra}chain two-particle 
scattering   and {\it inter}chain two-particle scattering processes, respectively, which we shortly describe below.   $\cal D$  symbolizes the 
measure of the path integral over the fermionic 
Grassmann variables.
In Figs.~1(a), 1(b) and 1(c), we show fundamental processes included in $S^{(1)}_{\perp}$, $S^{(2)}_{\para}$  and 
$S^{(2)}_{\perp}$, respectively. The interchain two-particle   processes included in $S^{(2)}_{\perp}$ are {\it generated} in the course of
the scaling.

The {\it intra}chain one-particle dispersion
is linearized at the Fermi points $\pm k_{F}$ and is given by  
$\varepsilon_{R}(\kpara)=v_{F}(\kpara-k_{F})$  and
$\varepsilon_{L}(\kpara)=v_{F}(-\kpara-k_{F})$ with $v_{F}$ being the Fermi velocity.
In both branches ($L$ and $R$) of bands,  the 
energy variables, $\varepsilon_{\nu  }$ ($\nu=R,L$ ), run over the region,
$-E/2<\varepsilon_{\nu }<E/2$, with
$E$ denoting the  bandwidth cutoff.\cite{Solyom}
The interchain one-particle hopping process [Fig.~1(a)] causes
a dispersion,  
$\varepsilon(\kperp)=-2t_{\perp}\cos\kperp$.
Momenta along and perpendicular to
the chain  are denoted by $k_{\para}$ and $k_{\perp}$,  respectively. 

At   half filling, the intrachain interaction generates the {\lq\lq normal\rq\rq} [Fig.~1(b-1)] and {\lq\lq umklapp\rq\rq} [Fig.~1(b-2)] scattering processes   with
 dimensionless scattering strengths $g_1$,  $g_2$ and $g_3$ 
of the backward, forward and umklapp scatterings, respectively.\cite{Solyom} The usual scattering strengths with a
 dimension of the interaction energy are $\pi v_{F}g_i$. The intrachain scattering strengths are related to the on-site and nearest-neighbor Coulomb repulsion, $U$ and $V$, as
$\pi v_{F}g_1=\pi v_{F}g_3=U-2V$ and $\pi v_{F}g_2=U+2V$ at  half filling only.

The 3rd order scaling equations for the intrachain two-particle processes [Fig.~1(b)] are given by\cite{Kimura75}
\begin{eqnarray}
{d g_{1}/ dl}&=&-{g_1}^2-{g_1}^3/2,\\
{d G/dl} &=&-{g_3}^2\left(1+G/2\right),\\
{d g_{3}/dl} &=&-g_{3}G\left(1+G/4\right)-{g_{3}}^{3}/4,
\end{eqnarray}
where $G=g_{1}-2g_{2}$.
The scaling parameter is defined by $l=\ln [E_0/ T]$ with $E_{0}$ and $T$ being the initial bandwidth cutoff and the absolute temperature, respectively.
The charge degrees of freedom in the intrachain system are governed by the combination $(G,g_3)$ with  flow lines $G^2-{g_3}^2={\rm const}$.
When the initial values, $G(0)$ and $g_3(0)$ satisfy the condition, $G(0)< \mid g_{3}(0) \mid $, the umklapp process becomes relevant
and (1) $\sim$ (3) give the non-trivial fixed point, $g_1^\ast=0$ and $\mid g_{3}^{\ast}\mid=-G^{\ast}=2$.
In this region, a Mott gap opens in the charge excitation spectrum and
 the antiferromagnetic (AF) correlation is the most dominant one (incipient instability) in the intrachain system.

The scaling equation for $t_{\perp}$ [Fig.~1(a)] is given by\cite{Kimura75,Bourbonnais93}
\begin{eqnarray}
d  \ln t_{\perp}/dl=1- \left({g_1}^2+{g_2}^2-{g_1}{g_2}+ {g_3}^2/2\right)/4.\label{eqn:tperp}
\end{eqnarray}
In Fig.~2, we show the diagrams which contribute to    the r.h.s of (\ref{eqn:tperp}). The renormalization of $t_\perp$ comes
only from the intrachain self-energy processes.
The non-trivial fixed point  gives
\beq
{d\ln t_{\perp}(l)/ dl}\stackrel{l\to\infty}{\longrightarrow}{1/ 4}.\label{eqn:FP}
\eeq
Thus for large $l$, $t_\perp$ grows as $t_\perp=t_\perp(0)e^{l/4}$. 
Consequently, in the present system, $ t_{\perp}$ is a {\it relevant} perturbation and always attains an order of  the initial bandwidth, $E_{0}$, 
at some crossover value of the scaling parameter, $l_{\rm cr}=\ln [E_{0}/ T_{\rm cr}]$, qualitatively defined by
\begin{eqnarray}
t_{\perp}(l_{\rm cr})=E_{0}.
\end{eqnarray}
$T_{\rm cr}$ corresponds to $T_{x1}$ which was   introduced by Bourbonnais and Caron.\cite{Bourbonnais91}
The system undergoes the one-particle crossover around $T_{\rm cr}$.

Here we note that    the similar scaling arguments give  
$
{d\ln t_{\perp}/ dl} \stackrel{l\to\infty}{\longrightarrow} 1-U^{2}/16\pi^2v_{F}^2 
$
for the case of a coupled non-half-filled Hubbard chain system,\cite{Bourbonnais91} and
$
{d\ln t_{\perp}/ dl} \stackrel{l\to\infty}{\longrightarrow} -{ U^{2}/ 32\pi^2v_{F}^2} -{7/8} 
$
for  the case of a coupled non-half-filled Hubbard ladder system.\cite{JKKY} 
In the former case, $t_{\perp}$ rapidly grows  for small $U$ and consequently the one-particle crossover always domintates the
two-particle crossover as far as $U$ is not extremely large.\cite{Bourbonnais91}
On the contrary, in the latter case, a spin gap opening in the intraladder system  so strongly reduces the growth of $t_{\perp}$ that 
 there   appears a region where the two-particle crossover dominates the one-particle crossover.\cite{JKKY}
We see that, in the case of the weakly-coupled half-filled  chain system, {\it $ t_{\perp}$ is    relevant but
 the growth of the intrachain umklapp process} ({\it Mott gap opening}) {\it strongly reduces the growth of $t_\perp$} and consequently
$t_\perp$   grows much more slowly than  in the weakly-coupled non-half-filled chain system. Consequently, as we shall show below, there appears a region
where the two-particle crossover dominates the one-particle crossover as in the coupled ladder system.

The action for the interchain two-particle  AF-interaction [Fig.~1(c)] is written as
\beq
S_{\perp\rm AF}^{(2)}&=&-{\pi v_{F}\over 4 }\sum_{Q}{ \cal V}_{\rm AF}
{\cal O}_{{\rm AF}}^{\ast}{\cal O}_{{\rm AF}}\non\\
&-&{\pi v_{F}\over 4 }\sum_{Q}{ \cal V}_{\rm um}\left[{\cal O}_{{\rm AF}}^{\ast}{\cal O}_{{\rm AF}}^{\ast}+{\cal O}_{{\rm AF}}{\cal O}_{{\rm AF}}\right].
\label{eqn:VW}
\eeq
The AF particle-hole field is defined as
${\cal O}_{\rm AF}(Q)$  $=T^{1/2}$ $\sum_{K,\sigma}$  $R_{\sigma}(K+Q){\vec \sigma}_{\sigma\sigma'}L_{\sigma'}(K)$,
 with  
$K=(k_{\para},k_{\perp},\ii \varepsilon_{l})$, $Q=(q_{\para},q_{\perp},\ii \omega_{n})$, $q_\para=2k_{F}=\pi$,
$\varepsilon_{l}=(2l+1)\pi T$ and $\omega_{n}=2n\pi T$ being  fermion and boson thermal frequencies, respectively. 
Of course, the interchain  processes contain  channels other than AF: charge-density-wave, singlet superconductivity and triplet superconductivity channels.
In this work we consider only the region where growth of  the AF correlation dominates the other channels.

Contributions to the lowest-order   scaling equations for ${\cal V}_{\rm AF}$ [Fig.~1(c-1)] and ${\cal V}_{\rm um}$ [Fig.~1(c-2)] are depicted in 
Figs.~3(a) and 3(b), respectively,   and are written as
\begin{eqnarray}
{d {\cal V}_{\rm AF}\over dl}&=&{1\over 2}{  t_{\perp}}^2\left({g_2}^2+4{g_3}^2\right)\cos q_{\perp}
\non\\
&+&{1\over 2}\left(g_2 V
_{\rm AF}+4g_3 {\cal V}_{\rm um}\right)
-{1\over 4}\left({{\cal V}_{\rm AF}}^2+4{{\cal V}_{\rm um}}^2\right),\label{eqn:V} \\
{d { \cal V}_{\rm um}\over dl}&=&2{  t_{\perp}}^2
{g_2}{g_3}\cos q_{\perp}\non\\
&+&2\left(g_2 {\cal V}_{\rm um}+g_3 {\cal V}_{\rm AF}\right)-{{\cal V}_{\rm AF}}{{\cal V}_{\rm um}}\label{eqn:W},
\end{eqnarray}
where we put $q_{\perp}=\pi$.
In Figs.3(a-1), 3(a-2) and 3(a-3), we show diagrams which contribute to the first, second and   third terms on the r.h.s of eq.(\ref{eqn:V}), which
play  separate roles.
Although,    ${ \cal V}_{\rm AF}={ \cal V}_{\rm um}=0$ at the initial step,
   the first term    generates a finite magnitude of 
${\cal V}_{\rm AF}$. 
At the intermediate step, the second term induces an exponential growth 
of ${\cal V}_{\rm AF}$.
At the final step, the third term  causes divergence of  ${\cal V}_{\rm AF}$ at a 
critical scaling parameter $l_{N}=\ln [E_{0}/ T_{N}]$ defined  by
\beq
{\cal V}_{\rm AF}(l_{N})=-\infty\label{eqn:defoflc2}.
\eeq
In Figs.3(b-1), 3(b-2) and 3(b-3), we show diagrams which contribute to the first, second and   third terms on the r.h.s of eq.(\ref{eqn:W}).
It follows from the structure of   eq.(\ref{eqn:W}) that  divergences of ${\cal V}_{\rm AF}$ and ${\cal V}_{\rm um}$ occur at the same $l_{N}$.

To see which of $T_{\rm cr}$ and $T_{N}$ is larger, we solve the coupled scaling equations (1) $\sim$ (4),  (8) and (9).
For the   initial conditions, we take 
\begin{eqnarray}
\left.\begin{array}{c}
U=4V=0.4\pi v_{F},\,\,\,\,
t_{\perp}(0) =t_{\perp0},\\
{\cal V}_{\rm AF}(0)=
{\cal V}_{\rm um}(0)=0.
\end{array}\right\}
\end{eqnarray}
In Figs.~4(a) and 4(b), we show the scaling flows of $t_{\perp}/E_{0}$, 
${\cal V}_{\rm AF}$ and ${\cal V}_{\rm um}$ for  
$ t_{\perp 0}=0.05E_0$ and 0.2$E_0$,
where  the vertical  lines show  locations of $l_{\rm cr}$ and $l_{N}$.
We see that, for $ t_{\perp 0}=0.05E_0$, $T_{\rm cr}<T_{N}$, while, for $ t_{\perp 0}=0.2E_0$, $T_{\rm cr}>T_{N}$.
In Fig.~4, we also show the scaling flows of the   stiffness of the intrachain charge excitation, $K_{\rho}=\sqrt{(2+G)/(2-G)}$.
Note again that the low-energy asymptotics of the isolated 1D electron system at  half filling is characterized by $K_{\rho}=0$, which corresponds to a fully opened Mott gap.

In Fig.~5, we show a phase diagram  spanned by $t_{\perp0}/E_0$ and $ T/E_{0}$.
Roughly speaking, we may regard increasing $  t_{\perp0}$ as increasing applied
 pressure. 
As the temperature  decreases,   the effectively isolated chain systems are gradually scaled from the Tomonaga-Luttinger liquid (TL) regime
to their low-energy asymptotics, the 1D  Mott insulator phase   characterized by $K_{\rho}=0$.
The gradual change of   darkness in the {\it incipient} 1D Mott  insulator  phase in Fig.~5 
schematically illustrates this situation.

There exists a crossover value of $t_{\perp}$: $  t_{\perp \rm cr}\sim 0.1 E_0$.
For $0< t_{\perp0}< t_{\perp \rm cr}$, as the temperature decreases, the system undergoes the crossover from the TL liquid to the incipient 1D Mott insulator and finally 
the two-particle crossover leads the system to the  phase transition into 
the AF state   at $T_{N}$. 
Dependence of $T_{N}$ on $t_\perp$ obtained here supports the result obtained by Bourbonnais\cite{Bourbonnais97} based on  the Stoner criteria for
the interchain AF coupling and
qualitatively  reproduces 
the experimentally observed pressure dependence of $T_{N}$.\cite{Jerome91}
In the real quasi-one-dimensional  (TMTTF)$_2$Br salts, there exists a small but finite $t_{\perp 0}$ even under ambient pressure. So it is appropriate to interpret that   
in our phase diagram, Fig.~5,  the region where  $t_\perp$ is very small is missing in reality. 
In this work, since we limited our attention only to a purely electronic system and neglected coupling of AF fluctuations to the $2k_{F}$ lattice distortions which becomes
important in the case of strong dimerization,
our view cannot cover the competition between  
the spin-Peierls  and AF phases\cite{Inagaki83} which is actually observed in the low effective pressure region of this series of salts.\cite{Pouget82}
To reproduce the experimentally observed temperature dependence of $T_{\rho}$ is also beyond our scope, since the feedback effects of $t_{\perp}$ on the
 intrachain system is not taken into account in the present scheme.

For $ t_{\perp \rm cr}<t_{\perp0}$, 
the system undergoes a crossover from the TL phase to  the Fermiology (FL) regime, where the interchain   propagation of a quasi-particle 
is coherent and  the 1D incipient instability is completely  lost.
   Based on the fact that, around  the crossover  scaling parameter, $l_{\rm cr}$,   the stiffness of
 the intrachain charge exciation is  
 far  from   $K_{\rho}=0$ corresponding to the fully opened Mott gap [see Fig.~4(b)], 
 we   interpret that the system undergoes the crossover from the TL phase 
directly to the FL regime around $T_{\rm cr}$ without visible Mott gap opening.

In the FL phase,  if the correlation is weak, the physical nature of
the system would be well understood in terms of the topology of the anisotropic 2D Fermi surface, 
where  the system adjusts  the SDW  vector to the {\it best nesting}   of the Fermi surface.\cite{Yamaji83}
In such a case the SDW transition temperature, $T_{\rm SDW}$, decreases with the increasing pressure, 
since the pressure tends to decrease the degree of the nesting.\cite{Yamaji83}

In Fig.~6, we show how the crossover value, $t_{\perp \rm cr}$,  
depends on the relative strength of the umklapp scattering, $g_{3}(0)/g_{1}(0)$,
for fixed $g_1(0)=U-2V$ and $g_2(0)=U+2V$ with $U=4V=0.4\pi v_{F}$.  
We see that {\it the crossover value, $\tilde t_{\perp \rm cr}$, always exists for 
a finite umklapp scattering.}
When the  umklapp scattering becomes less important, the AF phase  shrinks. 
This situation supports the   picture given by Emery {\it et al.}\cite{Emery82} that 
in  the TMTSF-series 
the  dimerization along the conducting stack is much weaker than that of the TMTTF-series and consequently  
  the umklapp process becomes less important even at high-temperature
region. Then  $g_3(0)$ becomes much smaller and very small $t_{\perp 0}$ is enough to locate the ground state in the FL reigme.

The  dimensional crossovers in the present case are quite analogous to those in  the weakly-coupled non-half-filled ladder system where a spin gap opening in the intraladder system
strongly reduces the growth of $t_{\perp}$.\cite{JKKY}
The  1D Mott insulator and the  AF phases of the present case correspond to the
spin-gap-metal (SGM) and the $d$-wave superconductivity (SCd) phase of the latter system, respectively.


We thank Professor K.Kanoda for valuable discussions and comments.
J.K was   supported by a Grant-in-Aid for Encouragement for Young Scientists   from the Ministry of Education, Science, Sports and Culture, Japan. 

\pagebreak
\noi
Fig.~1: A   zigzag line in (a) represents  the   {\it inter}chain one-particle hopping amplitude, $t_{\perp}$.
White and black circles represent  the {\it intra}chain two-particle {\lq\lq}normal{\rq\rq} (b-1) and {\lq\lq}umklapp{\rq\rq} (b-2) scatterings, respectively. 
White and black squares represent   the {\it inter}chain  tow-particle  antiferromagnetic (AF) interaction (c-1) and
 the umklapp process between electrons on different chains (c-2).
The 
solid and broken lines represent the propagators for the right-moving and 
left-moving electrons, respectively.
$i$ and $j$ denote   different chain indices.

\bigskip\noi
Fig.~2: Diagrams which contribute to the 
scaling of $t_\perp$.

\bigskip\noi
Fig.~3: Some typical diagrams which contribute to  the scaling of  ${\cal V}_{\rm AF}$ (a) and ${\cal V}_{\rm um}$ (b).

\bigskip\noi
Fig.~4: Scaling flows of $t_{\perp}/E_{0}$, 
${\cal V}_{\rm AF}$ and ${\cal V}_{\rm um}$ for  
$  t_{\perp 0}=0.05E_0$ (a) and $0.2E_0$ (b),
where  the vertical  lines show  locations of $l_{\rm cr}$ and $l_{N}$. 
A broken line represents the scaling flow of the stiffness of the intrachain charge excitation, $K_{\rho}$.

\bigskip\noi
Fig.~5:
Phase diagram of a weakly coupled chain system at   half filling.
The abbreviations are as follows: 
{\bf TL}= Tomonaga-Luttinger liquid,
{\bf 1DMott}=incipient one-dimensional Mott insulator,
{\bf FL}= Fermiology regime. Thick broken lines denote the crossover boundaries.

\bigskip\noi
Fig.~6:
Dependence of the crossover value $t_{\perp\rm cr}$ on the ratio $g_{3}(0)/g_{1}(0)$
for fixed $g_1(0)=U-2V$ and $g_2(0)=U+2V$ with $U=4V=0.4\pi v_{F}$.  
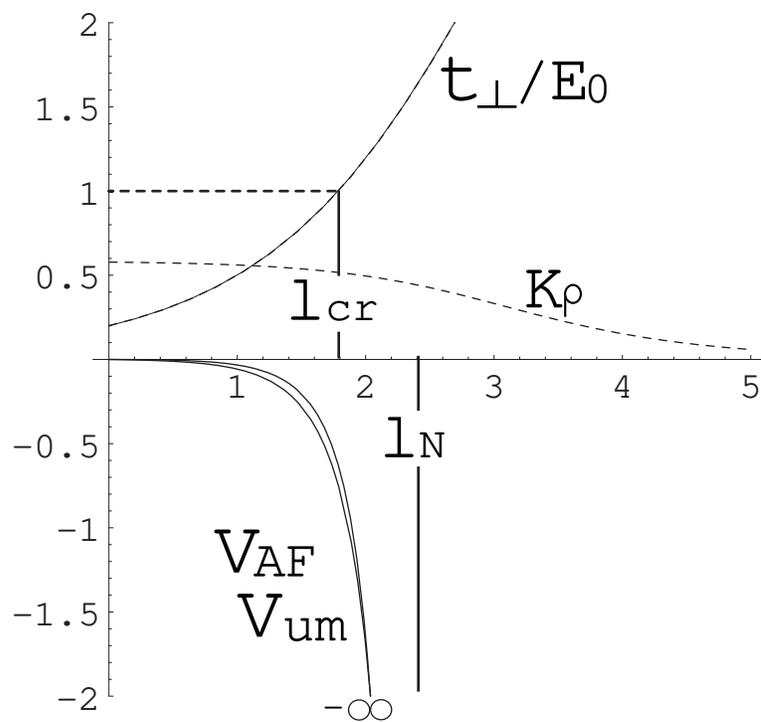
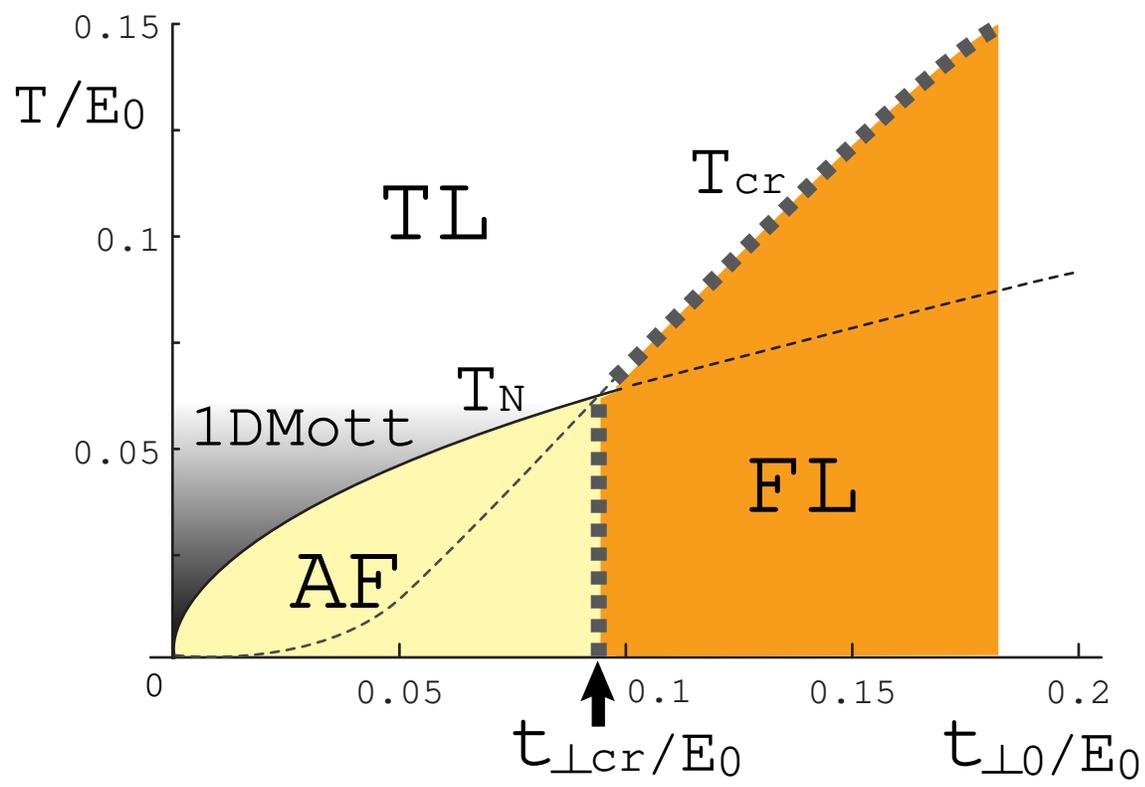
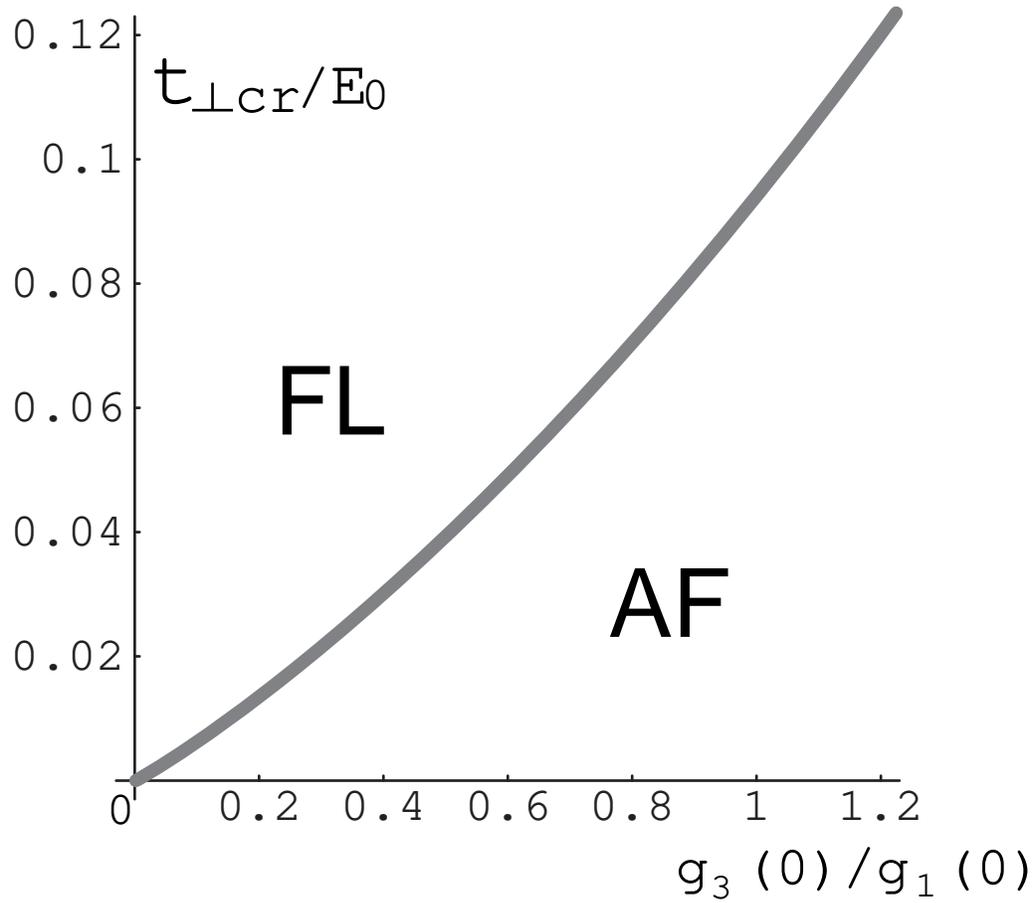
\end{document}